\documentclass[aps,pra,twocolumn,groupedaddress,showpacs]{revtex4-1}%\usepackage[latin1]{inputenc}
\usepackage[english]{babel}
\usepackage{setspace}  
\usepackage{graphicx}
\usepackage{amssymb}
\usepackage{mathtools}
\usepackage{amscd}
\usepackage{rotating}
\usepackage{color}
\usepackage{epstopdf}
\def\ve{\varepsilon}

\newcommand{\jqm}[1]{#1}
\newcommand{\jqd}[1]{}

\begin{document}

\title{Thermal versus Quantum Fluctuations of Optical Lattice Fermions}

\author{V. L. Campo, Jr.}
\affiliation{Departamento de F\'isica, Universidade Federal de S\~ao Carlos, Rodovia Washington Luiz, km 235, Caixa Postal 676, 13565-905, S\~ao Carlos, SP, Brazil}
\author{K. Capelle}
\affiliation{Universidade Federal do ABC, 09210-170, Santo Andr\'e, SP, Brazil}
\author{C. Hooley}
\affiliation{Scottish Universities Physics Alliance (SUPA), School of Physics and Astronomy, University of St Andrews, North Haugh, St Andrews, Fife KY16 9SS, U.K.}
\author{J. Quintanilla}\email{j.quintanilla@kent.ac.uk}
\affiliation{SEPnet and Hubbard Theory Consortium, School of Physical Sciences, University of Kent, Canterbury CT2 7NH, U.K.}
\affiliation{ISIS facility, STFC Rutherford Appleton Laboratory, Harwell Science and Innovation Campus, Didcot OX11 0QX, U.K.}
\author{V. W. Scarola}
\affiliation{Physics Department, Virginia Tech, Blacksburg, Virginia 24061, USA}

%\author{V. L. Campo, Jr.$^1$, K. Capelle$^2$, C. Hooley$^3$, J. Quintanilla$^{4,5}$, and V. W. Scarola$^6$}
%
%\affiliation{$\mbox{}^1$Departamento de F\'isica, Universidade Federal de S\~ao Carlos, Rodovia Washington Luiz, km 235, Caixa Postal 676, 13565-905, S\~ao Carlos, SP, Brazil}
%
%\affiliation{$\mbox{}^2$Universidade Federal do ABC, 09210-170, Santo Andr\'e, SP, Brazil}
%
%
%\affiliation{$\mbox{}^3$Scottish Universities Physics Alliance (SUPA), School of Physics and Astronomy, University of St Andrews, North Haugh, St Andrews, Fife KY16 9SS, U.K.}
%
%\affiliation{$\mbox{}^4$SEPnet and Hubbard Theory Consortium, School of Physical Sciences, University of Kent, Canterbury CT2 7NH, U.K.}
%
%\affiliation{$\mbox{}^5$ISIS facility, STFC Rutherford Appleton Laboratory, Harwell Science and Innovation Campus, Didcot OX11 0QX, U.K.}
%
%\affiliation{$\mbox{}^6$Physics Department, Virginia Tech, Blacksburg, Virginia 24061, USA}

\date{\today}

\begin{abstract}
We show that, for fermionic atoms in a one-dimensional optical lattice, the fraction of atoms in doubly occupied sites is a highly non-monotonic function of temperature.  We demonstrate that this property persists even in the presence of realistic harmonic confinement, and that it leads to a suppression of entropy at intermediate temperatures that offers a  route to adiabatic cooling.  Our interpretation of the suppression is that such intermediate temperatures are simultaneously too high for quantum coherence and too low for significant thermal excitation of double occupancy thus offering a clear indicator of the onset of quantum fluctuations.
\end{abstract}
\maketitle

\section{Introduction}

The international effort to emulate the behavior of correlated electrons in solids using ultracold atomic systems \cite{hofstetter2002,jaksch2005,bloch2008} is by now a familiar topic.  Experimental progress has been rapid, and the Mott insulating state of fermionic atoms in a 3D cubic optical lattice has been observed \cite{jordens2008,schneider2008}.  However, to explore the key open questions about the Hubbard model --- for example, how similar its phase diagram really is to that of the high-temperature superconductors \cite{lee2006} --- it is necessary to achieve a significantly lower entropy per particle than at present.  To be quantitative, the entropy per particle must drop below $k_B \ln 2$; this is the scale at which antiferromagnetic correlations begin to set in.

The role of theory in recent work on this problem has been vital.  On the one hand, experimental work is being conducted to help elucidate the properties of the Hubbard model at low temperatures.  On the other, the interpretation of the data taken in experiments often depends heavily on the theoretical understanding of the Hubbard model itself.  For example, one experiment \cite{schneider2008} measured the density profile of the cloud under varying amounts of harmonic confinement; the Mott insulator was identified by close comparison with theoretical predictions of the same.  In more recent experiments \cite{jordens2010}, measurements of the fraction of doubly occupied sites in the optical lattice, $d$, were compared with theory to show that these experiments are still dominated by thermal (rather than quantum) fluctuations, a topic we take up below.

As well as being relevant for interpretation of measurements on ultracold atomic systems, good theoretical understanding is crucial for the development of experimental cooling protocols.  Here also the behavior of the double-occupancy fraction as a function of temperature is important, as it has a direct relation to the entropic properties of the system.  For example, the Pomeranchuk-like cooling method outlined in \cite{werner2005} relies on an entropy enhancement, observable as a surprising suppression of double occupancy at intermediate temperatures.  Whether this effect is strong enough to be useful is thus a key question; recent numerical calculations tackling this issue in 3D lattices \cite{gorelik2010} are suggestive but results in 2D lattices \cite{paiva2010,khatami:2011} are mixed.  Highly accurate theoretical methods will be needed at low temperatures. 

Motivated by these requirements --- interpretation of data and design of cooling protocols --- we address in this work the following two questions.  First, how good is the double occupancy as a probe of the temperature?  And second, under what circumstances do we see the suppression of double occupancy at intermediate temperatures required for the adiabatic cooling protocol of Ref.~\cite{werner2005}?  We construct an accurate theoretical framework for treating low temperature fermions in one dimension.  We shall find that, in a 1D optical lattice, $d(T)$ is a non-monotonic function of temperature.  From the point of view of temperature measurement, this is a disadvantage; on the other hand, the resulting double-occupancy suppression indicates a large entropy enhancement, offering a clear route to adiabatic cooling.  We shall argue that the enhancement of double occupancy occurs for different reasons in different temperature regimes:\ it is driven by thermal fluctuations at high temperatures, but by quantum fluctuations at low temperatures.  We thus interpret the suppression in double occupancy at intermediate temperatures as occurring because the system is --- loosely speaking --- too hot for quantum coherence, but too cold for significant thermal excitation of double occupancy.  Finally, we shall show that these local effects can be measured \jqm{quantitatively} with a bulk observable even in the presence of realistic harmonic confinement.

Our method combines zero-temperature Bethe Ansatz (BA) studies with finite-temperature series expansions.  Recent implementations of BA \cite{snyder2010} compute observables such as the core compressibility \cite{scarola2009}.  Here, however we show that double occupancy alone indicates a large entropy enhancement, even though the 1D system is not a Fermi liquid.  By focusing on 1D Hubbard physics we demonstrate that exact calculations over the entire temperature range can be used as a platform to guide experiments in the construction of low temperature quantum states in optical lattice emulators.

%Body

\section{Model}

We consider the Hubbard model of cold fermionic atoms in a one-dimensional optical lattice:
\begin{equation}
  H=-  t\sum_{\langle i,j \rangle, \sigma}  c^{\dagger}_{i\sigma} c^{\phantom{\dagger}}_{j\sigma} + \text{H.c.}+U\sum_{j} n_{j\uparrow} n_{j\downarrow}-\sum_{j,\sigma} \ve_j n_{j\sigma}.
 \label{Hubbard}
\end{equation}
The indices $i$ and $j$ are integers labeling the sites of the lattice; the index $\sigma=\,\uparrow,\downarrow$ labels two different hyperfine states of the $N$ atoms in question, and we take both of these states to have equal total populations $N_{\uparrow}=N_{\downarrow}=N/2$.  The operator $c_{j\sigma}$ annihilates a fermion of species $\sigma$ on site $j$; the number operator $n_{j\sigma}$ is defined as $c^{\dagger}_{j\sigma} c_{j\sigma}$.

The first term in (\ref{Hubbard}) represents the quantum tunneling of atoms between neighboring sites of the lattice; the second the repulsion due to $s$-wave scattering when two atoms are present on the same site; and the third the single-atom site energy.  This last term may be site-dependent; for example, if the system were confined to a `box' of length $2L$ we would have $\epsilon_j=0$ for $ja\leq L$, $\epsilon_j = \infty$ otherwise. In the more realistic case of harmonic trapping, it would be given by 
\begin{equation}
\ve_j = ta^2j^2/L^2, \label{trappot}
\end{equation}
where $a$ is the lattice spacing and $L$ is a length scale related to the curvature of the trap.  The hopping parameter $t$ depends on the depth of the optical lattice; the on-site repulsion $U$ on the $s$-wave scattering length, which is tunable via Feshbach-resonance methods; and the on-site energies $\ve_j$ on the shape and amplitude of the trapping potential.
     
\begin{figure}[t] 
   \centerline{\includegraphics[width=3in]{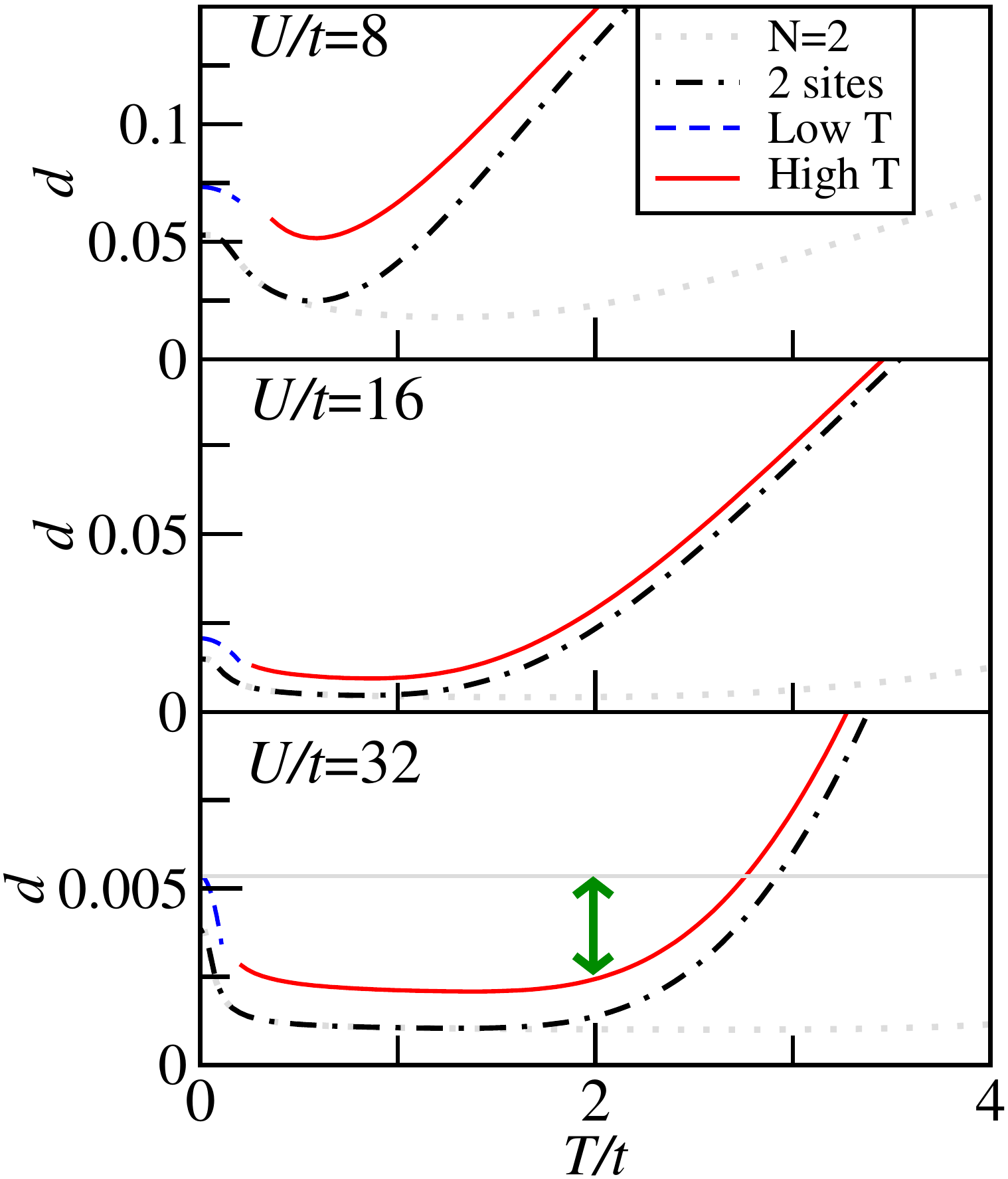}}
   \caption{Double-occupancy fraction versus temperature for the uniform 1D Hubbard model at half filling for several $U/t$.  The dashed line indicates the low-$T$ expansion (\ref{doublelowT}); the solid line is the fourth-order high-$T$ expansion; the dot-dashed line is the exact result for a two-site Hubbard model in the grand canonical ensemble; and the dotted line is the same but with the number of particles set to exactly two.  At low $T$, non-zero $d$ arises from coherent quantum fluctuations, but at intermediate temperatures this coherence is destroyed, leading to a suppression of $d(T)$ shown by the double-headed arrow.}
   \label{fig1}
\end{figure}
  
 \section{Results (homogeneous case)} 

We begin our analysis with an analytic study of the zero temperature double occupancy in the bulk limit (obtained by assuming a `box' potential as defined above and taking the thermodynamic limit $L,N \to \infty$, $N/L=$ constant) at half filling ($\mu=U/2$).
The double occupancy fraction, defined as the fraction of atoms in doubly occupied sites, is given by:
\begin{equation}
\label{doub_occ_fr}
d= (2/N) \left(\partial F /\partial U\right)_{T,N}=(2/N)\sum_{i} \langle n_i^{\uparrow}n_i^{\downarrow} \rangle,
\end{equation} 
where $F=E-TS$ is the free energy for entropy $S$ and $N$ is the total number of particles.   We compute the bulk double occupancy fraction at half filling from the exact solution of Lieb and Wu \cite{lieb1967}:
\begin{equation}
d_{\text{LW}}(U)=\int_{0}^{\infty}d\omega J_{0}(\omega)J_{1}(\omega) \text{sech}^{2}(\omega U/4).
\label{doublezeroT}
\end{equation}
Here, $J_n$ is the $n^{\rm th}$-order Bessel function.  Here and in the following we work in units of $t$ unless otherwise noted.  This expression shows that strong quantum fluctuations in the ground state induce finite double occupancy, even at $\langle n\rangle=1$ and large $U$.  

Weak thermal fluctuations compete with quantum fluctuations at low temperatures ($T\ll t$).  Using results from the quantum transfer matrix method \cite{klumper1996} we include thermal fluctuations to find the exact expression for the low $T$ double occupancy fraction:

\begin{equation}
d(T)\xrightarrow[T\ll t]{} d_{\text{LW}}(U)-C(2\pi/U)T^{2}+\mathcal{O}(T^{3}),
\label{doublelowT}
\end{equation}
where the function:
\begin{equation}
C(x)\equiv \frac{x^{2}}{12}\left[1-\frac{I_{0}(x)(I_{0}(x)+I_{2}(x))}{2I_{1}(x)^{2}}\right]
\end{equation}
is related to the unity central charge predicted by conformal field theory for the Heisenberg universality class \cite{klumper1996}.  $I$ is the $n^{\rm th}$-order modified Bessel function.  It is striking to note that $C>0$ for all $U$.  Thus weak thermal fluctuations counterintuitively \emph{lower} the double occupancy in the 1D Fermi Hubbard model.  In contrast, large thermal fluctuations dramatically increase double occupancy.  \\

\begin{figure}[t] 
\centerline{\includegraphics[width=3.4in]{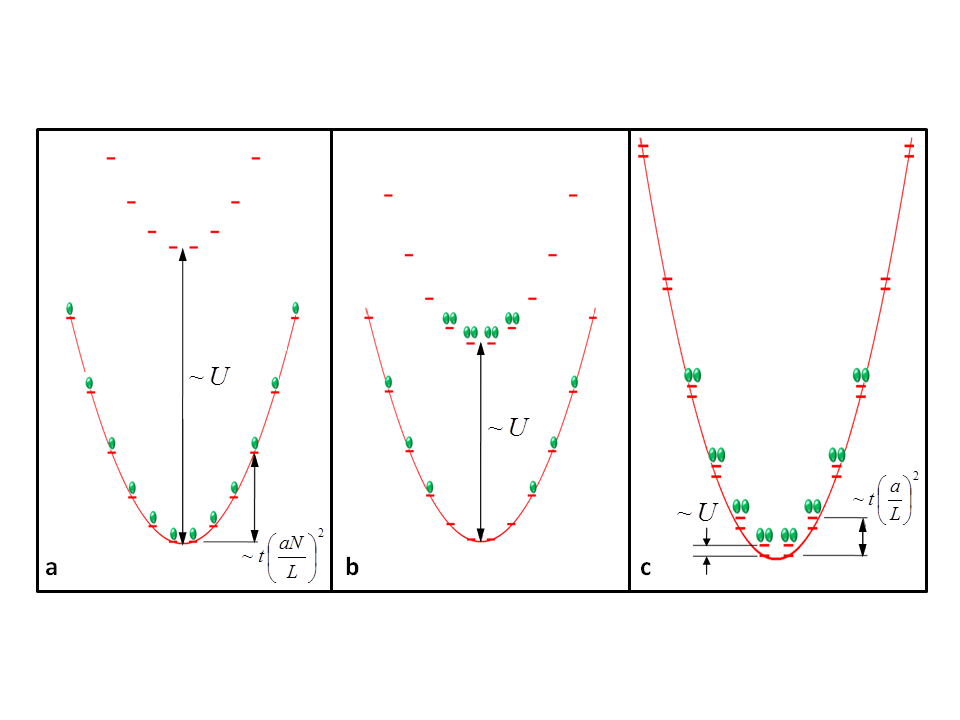}}
   \caption{Schematic of trap filling in three different regimes with decreasing $U$ from left to right.  Left:  Mott insulator with single occupancy, Center: Doubly occupied sites at the trap center flanked by edges with low filling, and Right: A band insulator dominated by doubly occupied sites. These sketches ignore both quantum and thermal fluctuations.}
   \label{fig2}
\end{figure}

We use a fourth order high temperature series expansion of the 1D Hubbard model to include large thermal fluctuations.  Expansion of the free energy in powers of $t/T$ about the atomic limit allows calculation of high temperature observables that compare well with finite temperature BA calculations \cite{takahashi2002}.  We compute the double occupancy fraction to fourth order.  For brevity, we present the equation only up to second order (for the full expression see the Appendix):
\begin{eqnarray}
&d&\hspace{-0.15cm}(T)\xrightarrow[T\gtrsim t^{2}/U]{}
-\left[
   TU+TU\cosh \left(\frac{U}{2 T}\right)\right]^{-1} +\frac{1-4U^{-2}}{e^{\frac{U}{2 T}}+1} \nonumber\\
&+& \tanh \left(U/4 T\right) \text{sech}^2\left(U/4 T\right)/4 T^2+2/U^{2}+\mathcal{O}(T^{-4})\nonumber
\end{eqnarray}
Figure 1 matches the low $T$ expansion (Eq.~\ref{doublelowT}) with the fourth order high $T$ expansion of $d$ within their respective regimes of convergence.  
Pronounced dips form when quantum and thermal fluctuations compete to dramatically enhance the entropy (The entropy can be obtained from the Maxwell relation $\left(\partial S/\partial U\right)_{T,N}=-(N/2)\left(\partial d/\partial T\right)_{U,N}$).  The finite value of $d$ at $T=0$ is due entirely to quantum fluctuations but the double arrow indicates the effect of thermal fluctuations at $T=2t$.  The related Pomeranchuk-like effect in higher dimensional lattices has been discussed in terms of Fermi liquid properties \cite{dare2000,werner2005} but it is well known that 1D Hubbard models exhibit non-Fermi liquid behavior.

We can understand the competition between quantum and thermal fluctuations in terms of a two site Hubbard model, which does not rely on Fermi liquid effects.  At half filling the two particle sector has the lowest energy at $T=0$ for $U>2t$.  
The ground state uses quantum fluctuations to lower its energy by hybridizing 
basis states. The singly occupied basis states correspond to the singlet $(\vert\uparrow,\downarrow\rangle-\vert\downarrow,\uparrow\rangle)/\sqrt{2}$ and the  
triplet $\{\vert \uparrow,\uparrow\rangle$, 
$(\vert \uparrow,\downarrow\rangle + \vert \downarrow,\uparrow\rangle)/\sqrt{2}$, $\vert \downarrow,\downarrow\rangle\}$ states. The singlet will mix with the doubly occupied states $\{\vert\uparrow\downarrow,0\rangle,\vert0,\uparrow\downarrow\rangle\}$ to lower its kinetic energy without paying a large penalty from the
interaction energy. The ground-state energy becomes $E_{0}=U/2-\sqrt{U^{2}/4+4t^{2}}\approx -4t^2/U$ for $t\ll U$.

At small but finite temperature, the three triplet states, all with energy $E_1=0$, allow thermal fluctuations to increase the entropy considerably (and therefore lower $F$) as we increase $T$.   However, these 3 excited states involve no double occupancy and therefore $d$ decreases.  As $T$ is increased further, however, even higher-energy states are populated. These other states, e.g., $\vert\uparrow\downarrow,0\rangle-\vert0,\uparrow\downarrow\rangle$,  involve double occupancy again, so the double occupancy goes up once more for large $T$. The non-monotonic behavior of double occupancy is closely-related to the separation of energy scales between the spin and charge degrees of freedom \cite{shiba1972,essler2005}.

\begin{figure}[t!] 
\centerline{\includegraphics[width=3.5in]{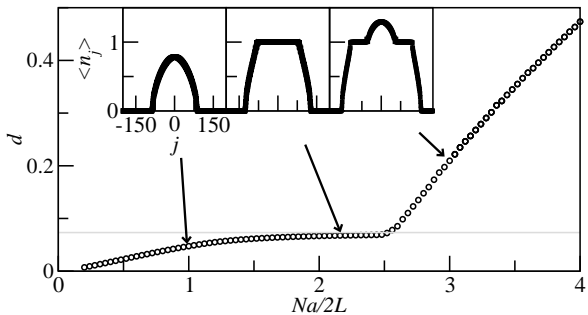} }
   \vspace{-0.8in}
   \caption{The circles indicate double occupancy fraction versus particle number in a trapped system ($L=50$) at $T=0$ for half filling at the trap center and $U/t=8$ computed using BA-LDA.  The solid line is the bulk value in the Mott insulating ground state (Eq.~\ref{doublezeroT}).  The insets, from left to right, show the corresponding density as function of position in the trap for $100$, $216$, and $300$ particles, respectively.}
   \label{fig3}
\end{figure}

Fig.~\ref{fig1} plots $d$ for the two-site Hubbard model for comparison with the thermodynamic limit.  The pronounced dip in $d$ corresponds to a finite $T$ entropy gain.
The remarkably close agreement between the expansions and the exact two-site results suggests that spinons and holons could be thought of in a localized picture.  The close agreement also suggests that entropy gain when increasing interaction at intermediate temperatures is a generic feature of few-particle bonding in non-frustrated Hubbard models.

\section{Results (harmonically trapped case)} 

We now address the competition between thermal and quantum fluctuations under realistic parabolic trapping. Let us first consider the situation in the absence of fluctuations, sketched in Fig.~\ref{fig2}. The left and right schematics show two scenarios where double occupancy should show a clear indication of the underlying trapped state.  For large $U$ (left) the deep Mott insulator regime corresponds to strong suppression of all doubly occupied sites.  The low $U$ regime (right) shows the band insulator, identifiable with a large number of doubly occupied sites.  The central schematic shows a mixture of both singly occupied sites at the edges and doubly occupied sites near the trap center. When fluctuations are turned on, the situation will be complicated further by the appearance of metallic regions and finite double occupancy at sites where $\left\langle n \right\rangle <2$.

At first it may appear that trapping will hamper efforts to distinguish between thermal and quantum fluctuations in observations of double occupancy. We will use the Bethe-Ansatz local-density approximation (BA-LDA)~\jqm{\cite{xianlong2006}} to show that bulk values are indeed observable in strongly inhomogeneous and relatively small trapped systems. Interestingly, local measurements~\cite{Ho_local} are not required to recover the bulk physics: a \emph{global} measurement carried out on the whole trapped system suffices. 

\jqd{BA-LDA is a highly accurate method related to the LDA used in solids but is distinct from "LDA" often applied in the optical lattice setting.  BA-LDA is done exclusively on the correlation energy, not the kinetic energy, which is obtained from the exact Bethe-Ansatz solution.  Furthermore, we implement BA-LDA through the Kohn-Sham scheme, which substantially improves the treatment of the kinetic term (in contrast with the Thomas-Fermi approximation).  As a result, BA-LDA shows excellent agreement with exact results where comparable \cite{xianlong2006}.}

We consider finite values of $L$ in Eq.~(\ref{trappot}) to study trapped systems with up to $N\sim 500$ particles with BA-LDA.  Fig.~\ref{fig3} plots the double occupancy fraction versus $Na/2L$, an effective filling in the trap in the thermodynamic limit $L,N \to \infty$, $Na/2L=$ constant \cite{rigol2003,campo2009}.  For low particle numbers, $Na/2L\lesssim 1$, the trap is entirely compressible due to edge effects.  Near $Na/2L\sim 2$ the central region forms a Mott insulator at the trap center.  At $T=0$ the system is dominated by quantum fluctuations manifest in the finite double occupancy even in a regime with $\langle n_{i} \rangle \leq 1$ for all $i$.  $d$ in the trap converges to $d_{\text{LW}}$ in the formation of a plateau.  Observation of this plateau indicates a strong Mott insulator with quantum fluctuations.  $d$ excludes edge effects and therefore allows a measurement of the bulk value of $d$, $d_{\text{LW}}$, even in an in-homogenous mesoscopic system.  Increasing $N$ further turns on double occupancy at the trap center leading to a pronounced cusp near $Na/2L\sim 2.5$.  

\begin{figure}[t] 
\centerline{\includegraphics[clip,width=3.8in]{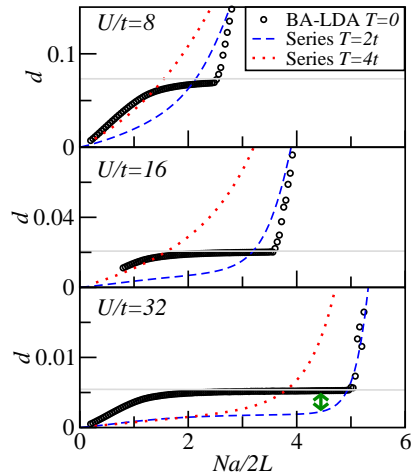}}
\vspace{-0.2in}
   \caption{The circles and solid line show the same as Fig.~\ref{fig3} but for three values of $U$.  The dashed (dotted) lines show results computed from an exact second order high temperature series expansion at $T=2t$ ($T=4t$).  The plateau at finite $d$ indicates strong quantum fluctuations.  The double arrow in the bottom panel indicates the observable effects of weak thermal fluctuations in a trap and corresponds to the double arrow in Fig.~\ref{fig1}. }
   \label{fig4}
\end{figure}

We use a $2^{\text{nd}}$ order high temperature series expansion to include finite temperatures in the trap.  The series from Ref.~\cite{scarola2009} is adapted to 1D.  The series we use for the trapped system is exact up to $\mathcal{O}((t/T)^{4})$.   Fig.~\ref{fig4} compares the second order series at finite $T$ and the BA-LDA at $T=0$.   From the top panel we see that strong thermal fluctuations tend to increase $d$ above the bulk value for the Mott plateau near $Na/2L\sim2.3$.  For larger $U$ the Mott gap suppresses thermal fluctuations.  In the bottom panels we see that thermal fluctuations tend to \emph{decrease} $d$ in the trap for $T\lesssim 2t$.  The double arrow line in the bottom panel shows that the effects of weak thermal fluctuations for $T=2t$ in a uniform system (Fig.~\ref{fig1}) are also observable in a trap.  We predict that even lower temperatures will tend to increase $d$ to its bulk value, $d_{\text{LW}}$ in the Mott regime.  Thus an increase of $d$ with decreasing $T$ demonstrates an observable capable of pinpointing a regime with dominant quantum fluctuations and low entropy per particle.  This regime can be used for adiabatic cooling and constructing higher dimensional optical lattices with low entropy.

\section{Connection to Adiabatic Cooling}

Our results complement a protocol for adiabatic cooling~\cite{werner2005}.  A Maxwell relation can be used to show that the thermal suppression of double occupancy implies a suppression of temperature with $U$~\cite{werner2005}.  Adiabatic cooling proposes to use changes in a tunable optical lattice parameter, $U$, to cross phase boundaries at fixed entropy.  The suppression of double occupancy that we find here implies that one dimensional systems offer a controlled platform for adiabatic cooling that can be used to systematically prepare low temperature systems in higher dimensions.  Lowering the optical lattice depth along one (or two) perpendicular directions in the lattice adds the term
$
-  t_{\perp}\sum_{\sigma,{\langle i,j \rangle}_{\perp}} c^{\dagger}_{i\sigma} c^{\phantom{\dagger}}_{j\sigma} + \text{H.c.}
$
to the Hamiltonian, increasing the dimensionality from 1D to 2D or 3D.
Thus by isentropically changing $t_{\perp}$, experiments will be able to identify and prepare higher dimensional optical lattice emulators from benchmark 1D configurations at entropies per particle where quantum fluctuations dominate, i.e., below $k_B \ln{2}$. Although this does not put
the system in a regime of low entropy per particle \emph{a priori}, adiabatic cooling may allow a 3D system to
cross the transition line from a paramagnetic to an antiferromagnetic phase.

\section{Conclusion}

We have studied the behavior of the fraction of atoms in doubly occupied sites, $d$, for the homogeneous and trapped Hubbard model. In the homogeneous system, $d$ shows a dip as function of temperature which signals the onset of antiferromagnetic correlations.  We also find that this important regime can be clearly identified even in harmonically trapped systems.  The double occupancy fraction increases with the number of particles in the trap, but remains flat when the trap has a large central plateau in its density profile.  The temperature behavior of $d$ in the harmonic trap with a large central plateau is similar to that of half-filled homogeneous systems and can be used to conclusively identify a regime with dominant quantum fluctuations and low entropy per particle.  Such an identification will set the stage for the preparation of states with antiferromagnetic correlations in optical lattices.

In preparation of this manuscript we recently became aware of related work in several different lattice geometries, Ref.~\cite{gorelik2011}, that also supports a similar entropy enhancement scenario to that put forward in Ref.~\cite{gorelik2010}a but beyond the DMFT approximation.

\section{Acknowledgements}

V.S.~acknowledges support from the Jeffress Memorial Trust (Grant No. J-992), AFOSR (FA9550-11-1-0313), and DARPA-YFA (Grant No. N66001-11-1-4122). K.C.~thanks Fapesp and CNPq. J.Q.~acknowledges funding from STFC, SEPnet, and the University of Kent Strategic Research Development Fund. V.L.C.J.~acknowledges support from the Brazilian agency CNPq.
\\

\section{Appendix}

Here we present an expression for the double occupancy fraction at half filling in units of $t=1$ up to fourth order in $t/T$ based on the free energy of 
Ref.~\cite{takahashi2002}:
\begin{widetext}
\begin{eqnarray}
&d&=\left[ 2 T^4 U^4
 (e^{\frac{U}{2 T}}+1)^5 \right]^{-1}e^{\frac{5 U}{4 T}} \Bigg [ 4 T \cosh \left(\frac{U}{4 T}\right) \Big \{3 T^3 U^4-T^3 
 \left(U^4-4 U^2+36\right) \sinh
   \left(\frac{U}{T}\right) \nonumber\\
   &+&T^2 U \left(T U^3+12\right) \cosh \left(\frac{U}{T}\right)   
 -4 T^2 \left(U^2-6\right) U 
   +U \left(4 T^2 (U^2 (T
   U-1)+9)-3 U^2\right) \cosh \left(\frac{U}{2 T}\right) \nonumber\\
   &+&2 T \left(-T^2 (U^4-4 U^2+36)+U^4-6 U^2\right) \sinh \left(\frac{U}{2
   T}\right)+6 U^3 \Big \} 
  +U^4 \left(\sinh \left(\frac{3 U}{4 T}\right)-11 \sinh \left(\frac{U}{4 T}\right)\right) \Bigg ]+\mathcal{O}(T^{-6}).
\end{eqnarray}
\end{widetext}

\end{document}